\documentclass[twocolumn,showpacs,preprintnumbers,amsmath,amssymb]{revtex4-1}
\usepackage{graphicx}
\usepackage{txfonts}
\usepackage{epstopdf}
\begin{document}

\title{Quantum Effects of Strain Influence on the Doping Energy in Semiconductors}

\author{Z. T. Wang$^1$}
\author{Shiyou Chen$^2$}
\author{X. M. Duan$^3$}
\author{D. Y. Sun$^1$}
\email{Email address: dysun@phy.ecnu.edu.cn}
\author{X. G. Gong$^4$}
\email{Email address: xggong@fudan.edu.cn}
\address{$^1$Department of Physics, East China Normal University,
Shanghai 200062, China}
\address{$^2$Laboratory of Polar Materials and Devices,
East China Normal University, Shanghai 200241, China}
\address{$^3$Department of Physics, Ningbo University, Ningbo 315211, China}
\address{$^4$Department  of Physics, Fudan University, Shanghai 200433, China }

\date{\today}

\begin{abstract}
Applying external strain is an efficient way to manipulate the site
preference of dopants in semiconductors, however, the validity of the previous
continuum elastic model for the strain influence on the doping
formation energy is still under debate. In this paper, by combining
quantum mechanical theoretical analysis and first-principles calculations, we show
that if the occupation change of different orbitals caused by the strain is negligible, the
continuum elastic model is valid, otherwise it will fail. Our theory is
confirmed by first-principles calculation of Mn-doped GaAs system. Moreover,
we show that under compressive strain the hole density, thus the Curie temperature $T_C$
can increase in Mn-doped spintronic materials.

\end{abstract}


\maketitle

Doping is an important technique to modify and tune material properties
in semiconductors.\cite{Schubert05} Usually, efficient doping requires that
the dopant stays at a specific site in the host crystal. For example,
to improve the Curie temperature $T_c$ for the spintronics application
of Mn-doped GaAs\cite{Ohno10,Die10}, Mn atoms should substitute at the Ga site (Mn$_{Ga}$) to create holes,
and the existence of Mn at the interstitial site (Mn$_{i}$) should be avoided because it
acts as donors which will kill holes that mediate the
ferromagnetism.\cite{yu02,Edmonds04,Dalpian06,Jungwirth06,Sato10} The site
preference of a dopant is determined by its formation energies at different
doping sites. Therefore, to increase Mn$_{Ga}$ and reduce Mn$_{i}$ concentration in GaAs,
one should reduced the impurity formation energy of Mn$_{Ga}$
and increase the impurity formation energy of Mn$_{i}$. To achieve this,
besides growing the sample at, e.g., Ga-poor condition, another approach could be
applying strain (stress), because the change of dopant formation energy under
strain is site dependent.\cite{Aziz97,Zhao99,Shu01,Walle03,Huang04,Aziz06,Huang06,Moriya93}

Many studies and progresses have been made to understand how dopant formation energy
at different sites is affected by strain. Zhang {\it et al.} have shown that an applied
pressure can stabilize the substitutional position of transition
metal impurities in Si.\cite{Zhang08} Theoretical calculations
by Chen {\it et al.} indicate that the formation energy of
self-interstitial atoms depends on strain.\cite{Chen10} More recently,
it has been shown by Zhu {\it et al.} using first-principles calculation that, for an
external hydrostatic or epitaxial strain, the change of the impurity formation
energy is monotonic, and it decreases if the external hydrostatic
strain is applied in the same direction as the volume change caused
by doping.\cite{Zhu10} In all these studies, classical continuum elastic
theory \cite{Dederichs78,Aziz97,Shu01,Zhu10} is used to explain experimental
and theoretical observations. In this theory, it assumes that the dopant
induces a stress on the
host crystal. When an external strain $\epsilon$ is applied, the formation energy
of the dopant, $\Delta E$, as a function of strain can be written as
\begin{equation}
\Delta E(\epsilon)=\Delta E_0 + V_0 P_D \epsilon~,
\end{equation}
where $\Delta E_0$ is the dopant formation energy without strain, V$_0$ is the equilibrium volume of the host.
$P_D$ is the pressure exerted on the host by the dopant. A similar relation between
$\Delta E$ and external stress $\sigma$ has also been proposed.\cite{Dederichs78,Aziz97}
Eq.~(1) clearly indicates that if the strain $\epsilon$ has the opposite sign as
$P_D$, the dopant formation energy will decrease
linearly as a function of $\epsilon$. The change will be large if $V_0P_D$ is large.

This simple formula has been widely used in the past\cite{Shu01,Huang04,Aziz06,Huang06}.
However, the origin of this formula is not well understood, especially
the validity of the linear relationship between the strain and the formation energy.\cite{Walle03,Chen10,Chan09,Ganster09}
Due to the importance of this formula for understanding the elastic theory,
the site preference of dopants under strain, and the
diffusion behavior between different sites, a detailed study and understanding
are needed.

In this paper, we find through quantum mechanical analysis and
first-principles calculation that the precondition for the continuum
elastic model and thus the linear dependence of the doping formation
energy on strain is that there is no discontinuity in the occupation
of the electron state when strain is applied.
If the occupation of the energy level is switched after applying the strain
due to band crossing in a quantum mechanical description, simple elastic model
described by Eq.~(1) will break down. This analysis is
supported by our direct calculation of the formation energy of
Mn-doped GaAs at different doping sites. We find that the change of the
formation energy of Mn$_i$ as a function of strain cannot fit to a linear equation because
under compressive strain electrons from conduction band derived donor level transfer
to the exposed unoccupied Mn $d$ level. Moreover, we show that a
compressive strain can enhance the concentration of free holes produced by
$Mn_{Ga}$ and suppress the formation of Mn$_i$, thus increase the hole density
and consequently the Curie temperature $T_c$.

We start our analysis by expanding the total energy of the host as a function of strain
in a Taylor series. According to the quantum theory, for a solid under small strain ($\epsilon$), the
energy up to the second order can be written as,
\begin{equation}
E(\epsilon)=E_0
+\epsilon\langle\Psi|\frac{\partial H(\epsilon)}{\partial\epsilon}|\Psi\rangle
+\frac{\epsilon^2}{2}\frac{\partial^2 \langle\Psi|H(\epsilon)|\Psi\rangle}{\partial\epsilon^2}~,
\end{equation}
where $E_0$ is the ground state energy of the host at $\epsilon = 0$,
$H$ and $\Psi$ are the Hamiltonian and wavefunction of the system,
respectively, and the derivatives are computed at $\epsilon = \frac{V-V_0}{V_0}=0$, i.e., at $V=V_0$.
Here, we assume volumetric strain.
Similar expression for other type of strain such as epitaxial or uniaxial strain can also be derived in a similar way.
The dopant formation energy is the energy difference between the
doped system and the undoped system.
The total energy of the doped system can also be expanded into a formula like Eq.~(1) except that
$E_0$ is replaced by $E_D$, the total energy of the doped system, and $\epsilon'=\frac{V-V_D}{V_D}$,
the strain with respect to the equilibrium volume of the doped system. By expanding the series of the
total energy of the doped system in terms of $\epsilon$ instead of $\epsilon'$ and subtracting that of the
host total energy, we see that
\begin{equation}
\Delta E(\epsilon)=\Delta E_0 + V_0 P_D \epsilon~ + \alpha\epsilon^2~,
\end{equation}
where, $\Delta E_0 = E_D - E_0$ is the dopant formation energy at equilibrium, and $P_D = -\frac{\partial E}{\partial V}|_{V=V_D}$
is the pressure exerted on the host by the dopant. The coefficient of the linear term $V_0P_D$
is proportional to the volume difference $\delta V_D=V_D-V_0$. It originates from the size difference between the dopant and the host elements. Because $\delta V_D$ is different for dopants
at different sites, the slope of the change of $\Delta E(\epsilon)$ will be different for different dopants, thus this property can be used to tune site preference of a dopant. The coefficient $\alpha$
of the quadratic term depends on the elastic constant difference between the doped and undoped system, therefore, it is negligible if the doping concentration is small, i.e., Eq.~(3) becomes Eq.~(1) at the impurity doping limit and if the applied strain is small.
This explains the origin of the continuum elastic model of Eq.~(1).

However, in our derivation above, we assume that the change of the Hamiltonian and wavefunction is a continuous function of $\epsilon$. This assumption is usually valid for semiconductors
because the valence band and conduction band will not cross each other under strain before
the metal-insulator transition. This situation also holds in most conventional
doped case when the defect levels (e.g., Si$_{Ga}$ donor level in GaAs) is derived from the band edge
state, so no band crossing is expected under strain. However, this assumption will not hold in a quantum systems where electrons can jump from one level to another level with very different atomic wavefunction characters when strain is applied, thus cause a discontinuous change in the Hamiltonian, i.e., the second term of Eq. 2.
This usually happens when the defect level has different origin from the host energy level, e.g., the $d$ orbital levels in a transition metal doped conventional semiconductors. From the analysis above, we can expect that after the charge transfer, the dopant changed from one state to another state with different size, therefore, $\delta V_D$ will change to $\delta V_D*$. In this case,
$V_0P_D$ itself is also dependent on $\epsilon$, thus the linear dependence will break down. If the charge transfer is gradual, then there will be a gradual transition region where the slope changes.

To demonstrate the above theory, we have calculated the formation
energy of Mn$_i$ doping in GaAs at two interstitial sites and
$Mn_{Ga}$ as a function of volumetric strain. Mn $3d$ band
is located near the band edge of GaAs, which provides a prototype example
to verify our theory above. The Mn-doped GaAs system is also a typical diluted magnetic semiconductor
for the development of the spintronic devices, thus, the calculated results also has significant
technical importance\cite{Ohno10,Die10}.

The first-principles calculations are carried out using Vienna {\it
ab initio} simulation program (VASP),\cite{Kresse96} based on
density functional theory with the projector augmented
wave (PAW) pseudopotential \cite{blochl94} and the generalized
gradient approximation (PBE-GGA)\cite{Perdew96} to the
exchange-correlation functional. The energy cut-off is set as 400
eV. The total energy and force on the atoms are converged within
$10^{-4}$ eV and $10^{-2}$ eV/\AA, respectively. A reciprocal space
k-point mesh of 4$\times$4$\times$4 for all the supercells is
employed. A (2$\times$2$\times$2) cubic supercell containing 64 host
atoms is adopted for all the defect calculations.
The optimized lattice constant is 5.75 $\AA$ for the zinc-blende
structure of pure GaAs, which is in good agreement with
previous studies.\cite{Wei04,Luo05}

In Fig. 1 we plot the strain influence on the calculated formation
energy of Mn doping in GaAs at four different sites: $Mn_{i\_As}$
interstitial surrounded by four As atoms, $Mn_{i\_Ga}$ interstitial
surrounded by four Ga atoms, Mn substitution at the Ga site,
$Mn_{Ga}$, and a defect complex $2Mn_{Ga}+ Mn_{i\_Ga}$.\cite{Jungwirth06,Mahadevan03,Baykov08}
The formation energy for a dopant at site $\alpha$ with the charge state q
is defined as
\begin{equation}\label{e-1}
\begin{split}
\Delta H(\alpha,q) = E(\alpha,q)-E(GaAs)+n_{Ga}[E(\rm{Ga})+\mu(Ga)]\\
-n_{Mn}[E(\rm{Mn})+\mu(Mn)]+q[\epsilon_{VBM}+E_F],
\end{split}
\end{equation}
where $E(\alpha,q)$ is the calculated total energy of the defect $\alpha$ in a supercell
in the charge state q, $E(GaAs)$ is the energy of pure
GaAs in the same supercell. $n_{i}$ is the number of Ga and Mn atoms that are
removed from the system during doping ($n_{Ga}=0$ and $n_{Mn}= -1$ for the interstitial
doping and $n_{Ga}=1$ and $n_{Mn}= -1$ for the substitutional defect). $E(\rm{Ga})$ and
$E(\rm{Mn})$ are the energy per atom of bulk Ga or Mn respectively,
and $\mu(\rm{Ga})$ and $\mu(\rm{Mn})$ are chemical potential of Ga and Mn, referenced to
the energy of bulk Ga and Mn, respectively. To improve p-type doping, Ga poor growth condition
with $\mu(\rm{Ga})$=-0.74 eV and $\mu(\rm{Mn})$=-0.61 eV are used in our
calculation.\cite{Mahadevan03} $\epsilon_{VBM}$ is the eigenenergy
of the valence band maximum (VBM) state of pure GaAs, and $E_{F}$ is the Fermi energy level
referenced to the VBM energy.

\begin{figure}
\scalebox{0.30}{\includegraphics{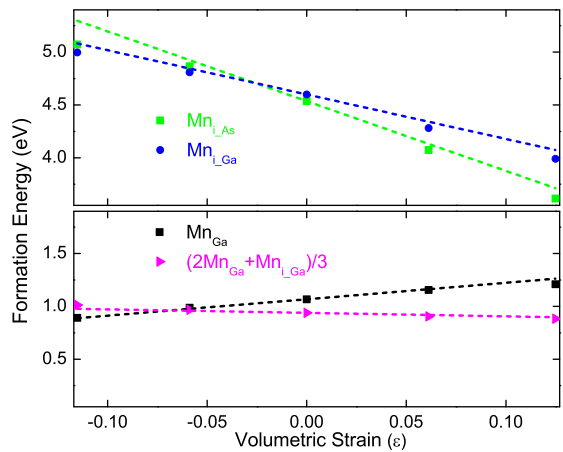}}
\caption{\label{fig1}(Color online) The calculated formation energies
as functions of the strain $\epsilon$ for Mn doping in GaAs at
different sites: $Mn_{i\_As}$ and $Mn_{i\_Ga}$ interstitials,
$Mn_{Ga}$ substitution, and $2Mn_{Ga}+Mn_{i\_Ga}$ complex per Mn dopant.
The predictions calculated from Eq.~(1) (straight lines) are plotted for comparison.}
\end{figure}

From Fig. 1, we can see two
obvious differences in the formation enthalpy change of substitution
and interstitial defects, (i) the formation energy increases with strain
for $Mn_{Ga}$ , whereas decreases for $Mn_{i\_As}$, $Mn_{i\_Ga}$ and the defect complex.
This can be explained according to the volume change caused by the
impurity formation. We find from our calculations that the volume is
decreased when a $Mn_{Ga}$ substitution is formed, and increased
when the interstitials are formed, so increasing $\epsilon$
will decrease the formation energy of interstitials and
increase that of substitutional dopant.\cite{Zhu10} (ii) the change is almost
linear for $Mn_{Ga}$ and the defect complex from $-0.115 < \epsilon < 0.125$,
in good agreement with that predicated by Eq. (1),
whereas the change is not linear for the two interstitial cases, $Mn_{i\_As}$
and $Mn_{i\_Ga}$, and is not agreement with what predicated by Eq. (1). In the following, we will explain why the nonlinearity exist
for the two donor defects $Mn_{i\_As}$ and $Mn_{i\_Ga}$ from the band structure of Mn-doped GaAs.

As a typical semiconductor, it is known that the top of valence band
of GaAs has mainly the bonding component of the p-p hybridization
between As and Ga, while the bottom of the conduction band mainly has
the antibonding component of the s-s hybridization between As and
Ga. When Mn, which has $d^5s^2$ atomic configuration, is introduced into
GaAs, the spin-up Mn 3d states are fully occupied whereas the
spin-down Mn 3d states are empty, producing impurity levels near the
conduction band edges.

For $Mn_{Ga}$ acceptor, it creates a hole at the top valence band,
so without strain $Mn_{Ga}$ has $d^{\uparrow5} + h$ configuration (Fig.2(left)).
Because both VBM and the localized Mn
$d$ orbital has low absolute deformation potential,\cite{li06} when
 strain is applied, the order of the occupied energy
level does not change. This explains why the linear relationship
between the formation energy and strain is kept for $Mn_{Ga}$, and can be well predicated by Eq. 1.

\begin{figure}
\scalebox{0.47}{\includegraphics{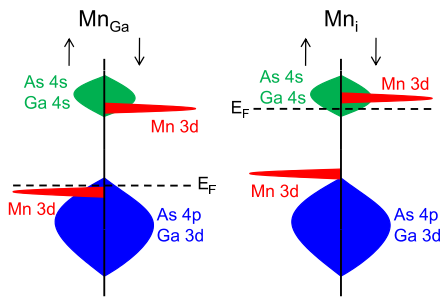}}
\caption{\label{fig2}(Color online) Schematic plot of the band
components of the Mn-doped GaAs and the position of the Fermi energies of
$Mn_{Ga}$ (left) and $Mn_{i\_As}$ (right).}
\end{figure}

In contrast to $Mn_{Ga}$, $Mn_{i}$ acts as a donor, generating two
electrons at the donor level near the conduction band maximum (CBM).
The spin-down Mn-3d band is only slightly higher than the donor level for $Mn_{i}$,
so it is not occupied when there is no strain.
Therefore, at $\epsilon=0$, $Mn_i$ has $d^{\uparrow5}+2e$
configuration (Fig.2(right)). When a compressive volumetric strain is applied to
the system, the antibonding conduction band as well as the derived
donor level with As 4s and Ga 4s character shifts upward in energy. On the
contrary, Mn 3d states are quite localized, so its energy level is
only weakly influenced by the strain. Due to the upward shift of the
donor level relative to the Mn-3d spin-down band, when $\epsilon <
-0.058$ electrons start to transfer from the CBM derived donor level
to the Mn-3d spin-down band, so the $Mn_i$ configuration changed to
$d^{\uparrow5}d^{\downarrow2}$ as the compressive strain is applied
to the $Mn_{i}$ doped system. The change of the atomic wavefunction
character from $s,p$ to $d$ make the simple linear relation of Eq.~(1) fail. Fig. 1
shows that under compressive strain, a linear line for $\epsilon >
0$ is replaced by another linear line for $\epsilon < -0.058$ with a
transition region between $-0.058 < \epsilon < 0.0$. The smaller
slope associated with the high compressive strain is because Mn in
$d^{\uparrow5}d^{\downarrow2}$ configuration at high compressive strain
is more localized, i.e., has a smaller size, than in the
$d^{\uparrow5}+2e$ configuration at zero or expansive strain.
Similarly, the smaller slope for $Mn_{i\_Ga}$ than for $Mn_{i\_As}$
is because there are more electrons around the anion atom As, so the
pressure exerted by Mn at an interstitial site next to As is larger
than that at an interstitial site next to Ga. All these are
consistent with our theory discussed above.

Our discussion above show that the reason that the formation energy
of $Mn_{i}$ does not follow the simple linear relation of
Eq.~(1) is because $Mn_{i}$ has two electrons near the conduction
band edge and the strain changes the electronic occupation of
different bands. Therefore, if we form the $Mn_{i\_Ga}+2Mn_{Ga}$
complex, so the two electrons from $Mn_{i\_Ga}$ passivate the holes of
$Mn_{Ga}$, then the strain in the range -0.115 to 0.125 will not
change the electronic occupation of the bands, thus the dependence
of its formation energy on the strain should be more linear. Indeed
our calculations confirmed this expectation, as shown in Fig. 1.
Therefore, our calculations demonstrated that in Mn-doped GaAs
system, whether the doping formation energy changes linearly with
the strain, i.e., the validity of the continuum elastic model,
depends on whether the strain changes the electronic occupation or
not. It should be mentioned that, although the current analysis are
demonstrated for the formation energy, the analysis also apply to
the energy differences and diffusion barriers between different
doping sites.

The above discussed dependence of formation energy on strain can
be used to tune doping properties for a specific applications.
For example, the experimentally
observed ferromagnetism of Mn-doped GaAs is mediated by the hole
produced by the $Mn_{Ga}$ acceptors.\cite{Jungwirth06,Dalpian06,Sato10}
However, as the hole density
increases and the Fermi energy shifts towards VBM, compensating
$Mn_{i}$ starts to form, which will lower hole density and thus
the Curie temperature $T_C$. Our calculated results in Fig. 1 suggest that
the relative ratio between $Mn_{Ga}$ and $Mn_{i}$ will increase significantly
when compressive strain is applied, therefore, the $T_C$ should increase
if compressive strain can be applied during the doping process.

In conclusion, the strain influence on the doping formation energy
at different sites in semiconductors is analyzed in terms of the
quantum mechanics theory and demonstrated by performing
first-principles calculation of Mn-doped GaAs system. We show that,
the validity of the continuum elastic model, i.e., whether the formation
energy is a linear function of the applied strain, depends on
whether the occupation of the difference electronic bands is changed or not.
If the occupation change caused by the strain is negligible, the
formation energy is linearly dependent on the strain, while if the
occupation change is significant, the linear relation, thus the continuum
elastic model will fail. Our study clarifies the previous
confusion about this linear relationship, and provides an easy way for
predicting the strain influence on the doping site in
semiconductors. The calculation also shows that the Fermi energy
pinning level of Mn-doped GaAs can be shift to close to VBM through
applying compressive strain to the system, which can increases the
ferromagnetic Curie temperature.

This research is supported by the Natural Science Foundation of
China and Shanghai, the Fundamental Research Funds for the Central
Universities,  Shuguang and Innovation Program of Shanghai Education
Committee. The computation is performed in the Supercomputer
Center of Shanghai and the Supercomputer Center of ECNU.

\end{document}